\documentclass[12pt]{iopart}
\usepackage[dvips]{graphicx}
\usepackage{setstack}

\begin{document}
\letter{The ferro/antiferromagnetic $q$-state Potts model}
\author{D P Foster and C G\'erard}
\address{Laboratoire de Physique Th\'eorique et
Mod\'elisation (CNRS ESA 8089)\\
Universit\'e de Cergy-Pontoise,
5 Mail Gay-Lussac\\ 95035 Cergy-Pontoise Cedex, France}

\begin{abstract}
The critical properties of the mixed ferro/antiferromagnetic
$q$-state Potts model on the square lattice 
are investigated using the numerical
transfer matrix technique. The transition temperature is found to
be substantially lower than previously found for $q=3$. It is
conjectured that there is no transition for $q>3$, in
contradiction with previous results.
\end{abstract}



\maketitle

In this letter we investigate the anisotropic $q$-state Potts
model on the square lattice\cite{potts,wu}. This model is defined
by the Hamiltonian
\begin{equation}\label{hamil}
{\cal H}=-\sum_{x,y}(J_x \delta(\sigma_{x,y}-\sigma_{x+1,y})+J_y
\delta(\sigma_{x,y}-\sigma_{x,y+1}))
\end{equation}
where the variables $\sigma_{x,y}$ may take one of $q$ distinct
values, and $(x,y)$ runs over the coordinates of the sites of the
lattice.

Much is known in the cases when $J_x$ and $J_y$ are of the same
sign\cite{wu,baxter}, corresponding to the ferromagnetic Potts
model ($J_x>0$, $J_y>0$) and the antiferromagnetic Potts model
($J_x<0$, $J_y<0$)\cite{afpotts}. Much less is known in the mixed
interaction case, $J_x>0$ and $J_y<0$. This will be the object of
this letter.

There have been a number of attempts to study the mixed
interaction model, particularly for the case
$q=3$\cite{kswu,ttt,selke,yasumura}. Kinzel, Selke and
Wu\cite{kswu} obtained a candidate critical line by applying a one
step Migdal-Kadanoff procedure, thereby mapping the model onto
the ferromagnetic Potts model for which the exact expression for the
critical line is known. This gave the critical line
as
\begin{equation}\label{kinzline}
(1+\exp(\beta J_x))(1-\exp(\beta J_y))=q.
\end{equation}
Spurred on by the fact that this line is exact for the exactly
known $q=2$ case, it was suggested that perhaps
this equation may remain exact for other values of $q$. They also
gave numerical results from Monte Carlo simulations which suggest
a transition line at lower temperatures than given by
\eref{kinzline}. The transition line was estimated, however, by
extrapolating the inverse correlation lengths by hand to zero for
lattice sizes less than about 60.

The possibility that \eref{kinzline} may correspond to an exact
result was refuted by the observation that, except for $q=2$, this
line does not remain invariant under the symmetries of the
model\cite{ttt}. The symmetries of the model may be used to
construct a symmetry group\cite{maillard}. The critical line must
remain invariant  under the action of these symmetries. Using the
group of symmetries of the model, Truong\cite{ttt} constructed
extended duality relations which he used to look for candidate
critical lines. For the 3-state Potts model this gave:
\begin{equation}\label{tttline}
\exp(\beta (J_x+J_y))+2\exp(\beta J_x)-\exp(\beta J_y)+1=0.
\end{equation}
Interestingly, this line has the same functional form as the
ferromagnetic transition line but with $J_y\to -J_y$. This line is
indeed invariant under all the symmetries of the model, but does
not seem to correspond to the transition line when compared with
numerical results of Kinzel \etal\cite{kswu,selke},
Yasumura\cite{yasumura} and of the present work. This leaves the
interesting question: Does this very special line have a
corresponding physical meaning? This question remains open.

It has been conjectured that the transition for $q=3$ is of
Kosterlitz-Thouless type. This conjecture is based on the
free-fermion approximation applied to a related
clock-model\cite{ostlund}, and supported by numerical results on a
related one dimensional quantum Potts model\cite{herrmart}. In
this letter we present convincing evidence that the transition for
$q=3$ is indeed of the Kosterlitz-Thouless type. 
The critical exponent $\nu$ is given
as a function of $q$ and shown to diverge at $q=3$. The critical
temperatures are also given, as a function of $q$, for the case
$J_x=-J_y>0$. No transition is found for $q>3$. This work provides
a more accurate estimate of the critical line for $q=3$, which is
substantially lower than previous
estimates\cite{kswu,selke,yasumura}.

The numerical results were obtained using the numerical Transfer
Matrix Method coupled with phenomenological
renormalisation\cite{nightingale}. The method consists of
expressing the partition function of an infinite strip of finite
width $L$ in terms of a matrix product. Imposing periodic boundary
conditions in the transfer direction (along the direction taken to
be infinite), the partition function for the strip may be written
\begin{equation}
{\cal Z}_L = \lim_{N\to \infty} \Tr T^{N}.
\end{equation}
The thermodynamic limit corresponds to the limit $L\to \infty$.
For integer values of $q>1$ it is possible to write the transfer
matrices directly in terms of spins. Full details are given in
Foster, G\'erard and Puha\cite{fgi}. For non-integer values of $q$
it is necessary to use the Kasteleyn-Fortuin mapping\cite{kf},
which gives the Potts model in terms of bond-clusters on the
lattice. Full details on the construction of the transfer
matrices for this case are given in Bl\"ote and
Nightingale\cite{blnight}.

It may
be shown that the correlation length in the transfer direction is given by
\begin{equation}
\xi_L=\log\left(\frac{\lambda_0}{\lambda_1}\right).
\end{equation}
Assuming scale invariance at the critical point, finite size estimates
of the critical temperature may be identified with solutions of the
equation
\begin{equation}\label{phenom}
\frac{\xi_L(T^*_{L,L^{\prime}})}{L}=\frac{\xi_{L^{\prime}}
(T^*_{L,L^{\prime}})}{L^{\prime}}.
\end{equation}
The critical exponent $\nu$ may be estimated at the fixed points,
solutions of \eref{phenom}, by
\begin{equation}
\frac{1}{\nu_{L,L^{\prime}}}=\frac{\log\left(\frac{{\rm d}\xi_L}{{\rm
d}T}/\frac{{\rm d}\xi_{L^{\prime}}}{{\rm d}T}\right)}{
\log\left(\frac{L}{L^{\prime}}\right)}-1
\end{equation}

\begin{figure}
\caption{The critical temperature as a function of $q$ for a transfer
direction along (A) the $x$-direction and (B) the
$y$-direction. Periodic boundary conditions are taken in both
directions.}\label{crit-temp}
\begin{center}

\vspace{.3cm}

\includegraphics[width=12cm]{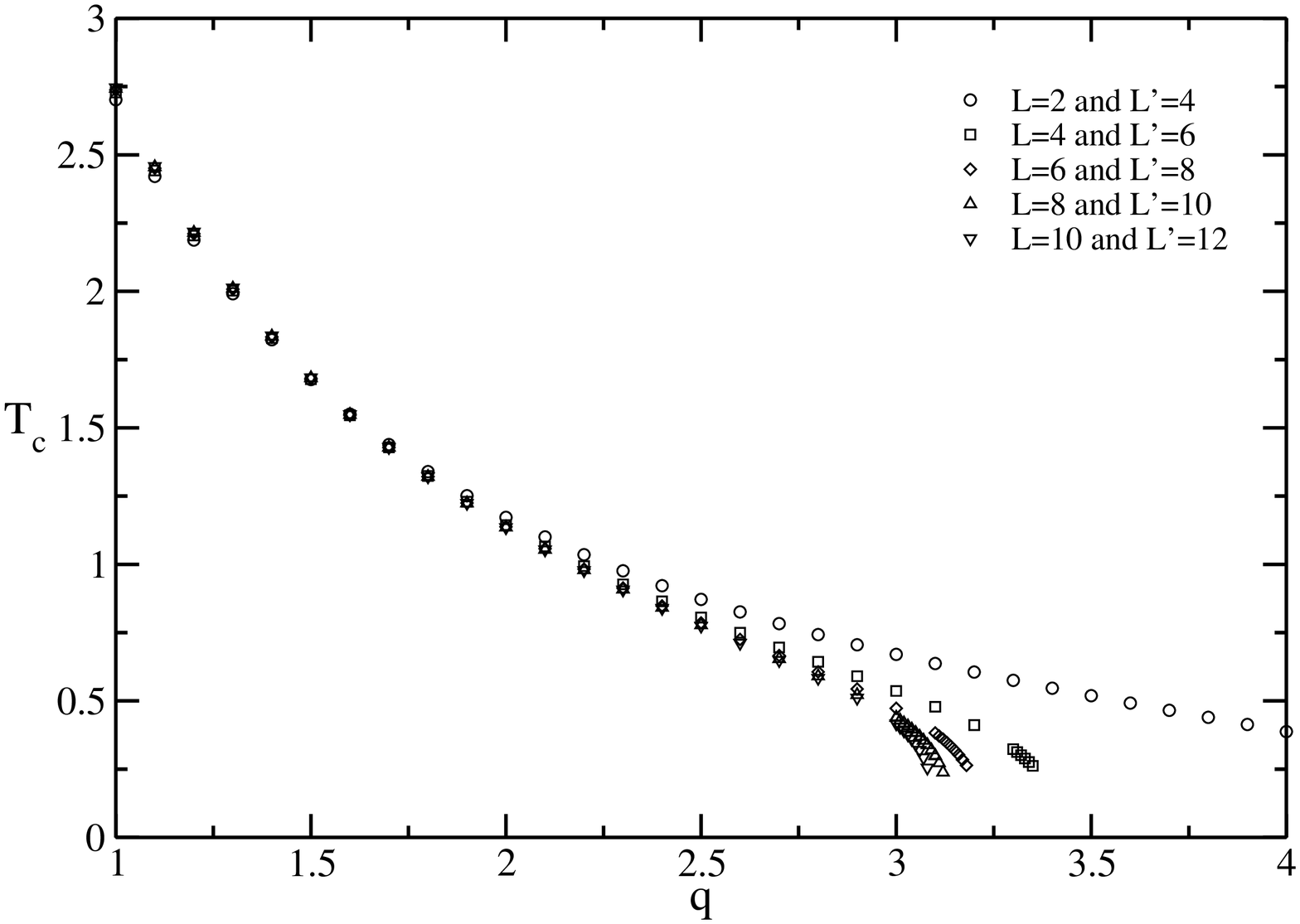}

\vspace{.3cm}

(A)

\vspace{.3cm}

\includegraphics[width=12cm]{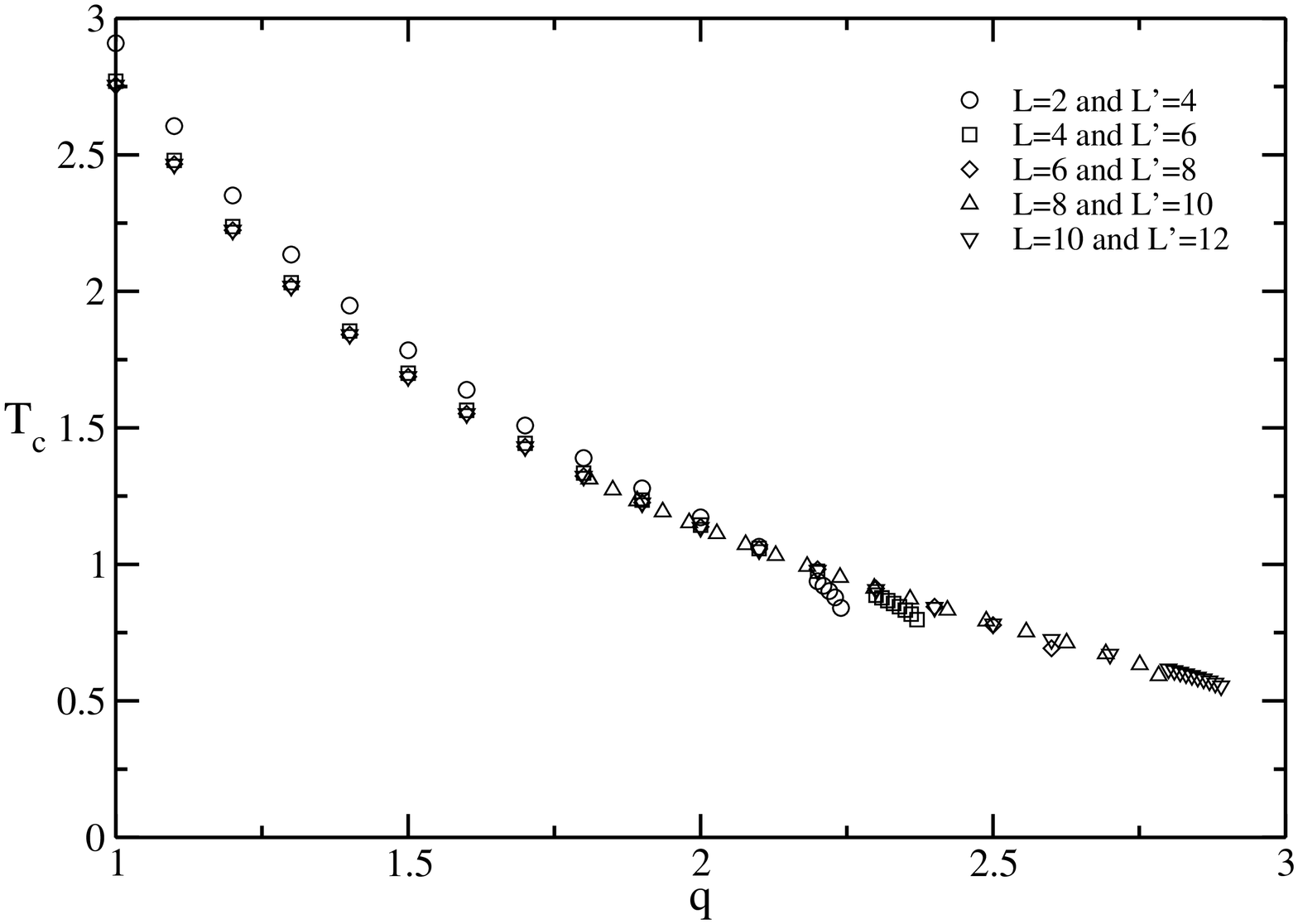}

\vspace{.3cm}

(B)
\end{center}
\end{figure}

\begin{figure}
\caption{The values of $q_{\rm max}$, the largest values of $q$ for
which an estimate of a critical temperature could be found, as a
function of $1/L$. The upper line corresponds to the transfer
direction in the $x$-direction, while the lower line of points
corresponds to the transfer direction in the $y$-direction. The point
$q=3$ and $1/L=0$ is shown for reference.}\label{qmax}
\begin{center}

\vspace{.3cm}

\includegraphics[width=12cm]{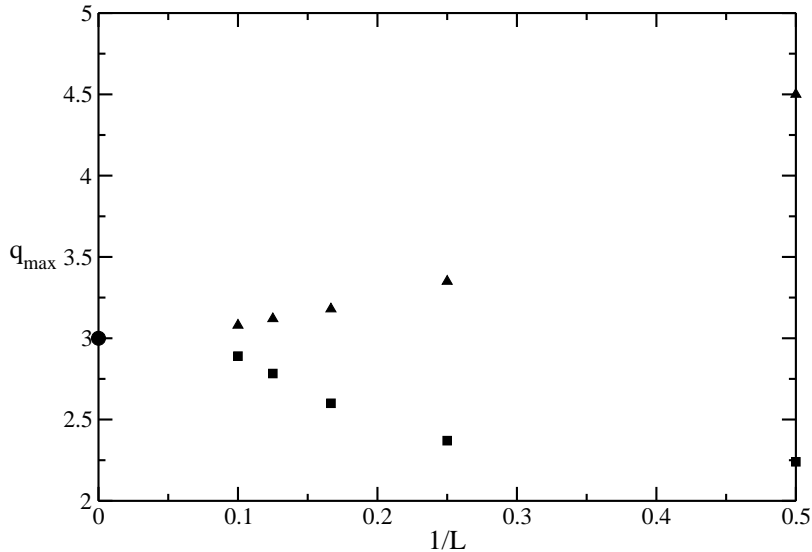}

\vspace{.3cm}
\end{center}
\end{figure}

Applying these results to the case $J_x=-J_y>0$ gives the results
shown in \fref{crit-temp} for the critical temperature as a
function of $q$. In \fref{crit-temp}~(A) the transfer direction is
taken in the $x$-direction, i.e. parallel to the ferromagnetic
interactions, and in \fref{crit-temp}~(B) the transfer direction
is taken in the $y$-direction, i.e. parallel to the
antiferromagnetic interactions. In the first case the largest
value of $q$ for which a solution was found, $q_{\rm max}(L)$,
decreases with $L$. On the other hand, $q_{\rm max}(L)$ increases
with $L$ in the second. This indicates a value of $q$, $q_{\rm
max}$, such that there exists a critical phase transition for
$q\leq q_{\rm max}$, but not for $q> q_{\rm max}$.
Figure~\ref{qmax} shows the values of $q_{\rm max}$ found in the
two cases. 
The curves both extrapolate plausibly to $q_{\rm max}=3$ in the
thermodynamic limit. Using the upper line in \fref{qmax} to
extrapolate, we qive $q_{\rm max}=3.00\pm0.003$.
This is in contradiction with Yasumura\cite{yasumura}, who gives
non-zero values of $T_c$ even for $q>3$.

\begin{figure}
\caption{Finite size estimates of $\nu$ as a function of $q$.}\label{nuest}
\begin{center}

\vspace{.3cm}

\includegraphics[width=12cm]{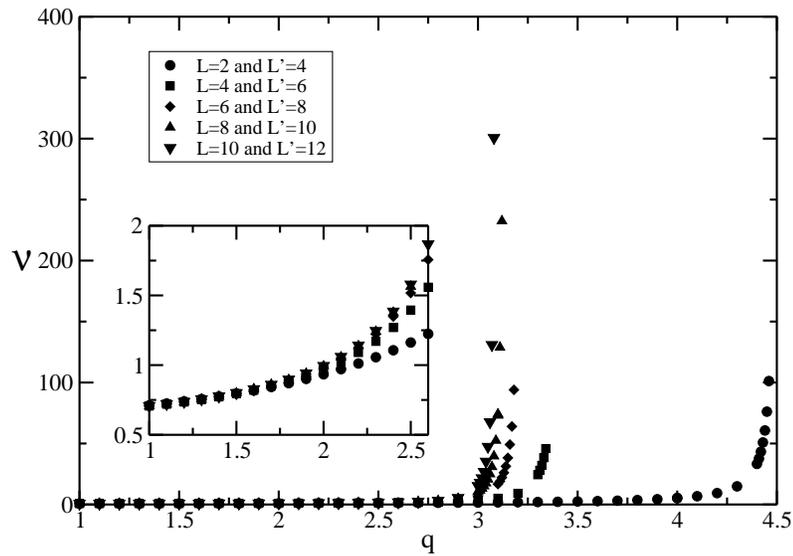}

\vspace{.3cm}
\end{center}
\end{figure}

The finite-size estimates of $\nu$ are shown in \fref{nuest} for a
transfer direction in the $x$-direction. They converge well to $\nu=1$
for $q=2$, and diverge for $q=3$. \Tref{critest}
shows the
finite-size estimates of $T_c$
for selected values of $q$. The estimates of $T_c$ are extrapolated
using a Burlich and Stoer algorithm for $q=1, 1.5, 2$ and 2.8, but had to be
extrapolated graphically for $q=3$.

\begin{figure}
\caption{Phase diagram for $q=3$. The cross shows the estimated
critical temperature of Kinzel \etal for $J_y/J_x=-1$, 
along with error bar. The star
shows the extrapolation of our estimated critical temperatures, also
for $J_y/J_x=-1$, with
error bar.}\label{phase}
\begin{center}

\includegraphics[width=12cm]{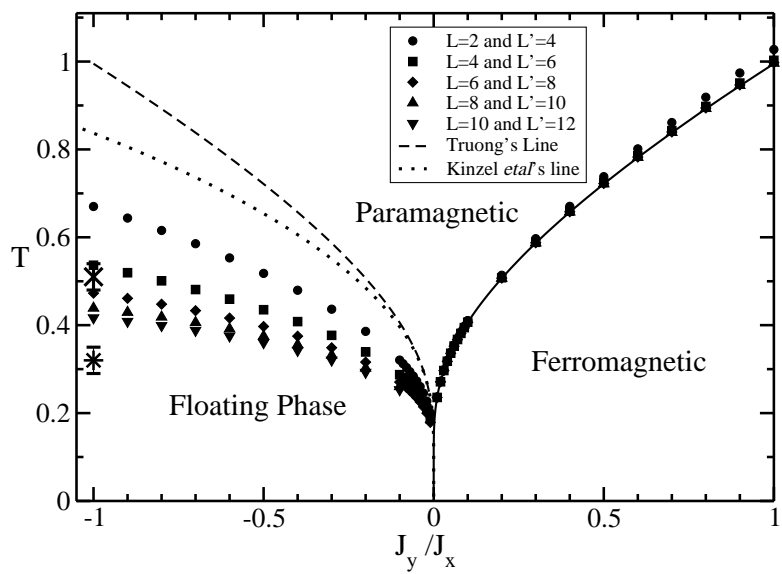}
\end{center}
\end{figure}

The phase diagram for $q=3$ is shown in \fref{phase}. The critical
line estimates obtained are compared with the exact line for
$J_y/J_x>0$, and shown to converge well, and are compared to both
proposed critical lines. The convergence is much slower for
$J_y/J_x<0$, but clearly rules out both proposed analytic
results\cite{kswu,ttt} and the numerical
results\cite{kswu,selke,yasumura}.

\begin{table}
\caption{Estimates of $T_c$ for selected values of $q$. The transfer
direction is taken in the $x$ direction.}\label{critest}
\begin{indented}
\item[]\begin{tabular}{@{}llllll}
\br
$L/L^{\prime}$ & $q=1$ & $q=1.5$ & $q=2$  & $q=2.8$ & $q=3$ \\
\mr
2/4 & 2.701840  & 1.677124 &   1.172009 & 0.742794 & 0.670151 \\
4/6 &  2.728013 &  1.678307  & 1.143901 &  0.642965 & 0.536430  \\
6/8 &  2.737827 & 1.680590  &1.137474  & 0.606052  & 0.472832\\
8/10 & 2.742918 & 1.682553 & 1.135844  & 0.590809  & 0.438814 \\
10/12 & 2.745847 & 1.683806 & 1.135255 & 0.583354  & 0.416617 \\
\mr
$\infty$  & $2.75\pm0.01$   & $1.69\pm0.01$  & $1.13\pm0.01$ & $0.57\pm0.01$
& $0.32\pm0.03$  
\\\br
\end{tabular}
\end{indented}
\end{table}

In summary, in this letter we have addressed  numerically the
nature of the criticality in the only sector of the anisotropic
Potts model for which exact solutions do not as yet exist: the
ferro/antiferromagnetic Potts model. We have given critical
temperature estimates and estimates of $\nu$ as a function $q$ for
$J_y/J_x=-1$. We find convincing evidence that there exists a
$q_{\rm max}$ around $q=3$ above which there is no transition. At
$q_{\rm max}$ the value of $\nu$ is diverging, consistent with a
Kosterlitz-Thouless type transition. This type of
transition has already been conjectured for $q=3$ using the
free-fermion approximation\cite{ostlund}. We conjecture, based on
this and our numerical data, that $q_{\rm max}=3$. This picture is
analogous with what happens in the $O(n)$ spin models in two
dimensions, 
where there
is a transition for $-2\leq n \leq 2$, the transition at $n=2$
corresponding to the Kosterlitz-Thouless transition\cite{kt}.

\ack

We would like to thank T T Truong for helpful discussions.

\end{document}